%% file: wnj.tex
\documentclass[aps,prl,twocolumn,fleqn,groupedaddress,nofootinbib,preprintnumbers,nobalancelastpage]{revtex4-1}
\pdfoutput=1
\input{defs}

\usepackage[pdfborder={0 0 0}]{hyperref}
\hypersetup{
  pdfauthor={Stefan Hoeche, Frank Krauss, Marek Schoenherr, Frank Siegert},
  pdftitle={W+n-jet predictions with MC@NLO in Sherpa},
  pdfkeywords={QCD, NLO, Parton Shower, MC@NLO}
}

\begin{document}

\preprint{SLAC-PUB 14859}
\preprint{IPPP/12/03}
\preprint{DCPT/12/06}
\preprint{LPN12-026}
\preprint{MCNET-12-01}
\preprint{FR-PHENO-2012-001}

\title{$W$+$n$-jet predictions at the Large Hadron Collider\\
  at next-to-leading order matched with a parton shower}

\author{Stefan~H{\"o}che}
\email{shoeche@slac.stanford.edu}
\affiliation{SLAC National Accelerator Laboratory, Menlo Park, CA 94025, USA}

\author{Frank~Krauss}
\email{frank.krauss@durham.ac.uk}
\author{Marek~Sch{\"o}nherr}
\email{marek.schoenherr@durham.ac.uk}
\affiliation{Institute for Particle Physics Phenomenology, Durham University, Durham DH1 3LE, UK}

\author{Frank~Siegert}
\email{frank.siegert@cern.ch}
\affiliation{Physikalisches Institut, Albert-Ludwigs-Universit{\"a}t Freiburg, D-79104 Freiburg, Germany}

\begin{abstract}
  For the first time, differential cross sections for the production of 
  $W$-bosons in conjunction with up to three jets, computed at next-to
  leading order in QCD and including parton shower corrections, are presented 
  and compared to recent experimental data from the Large Hadron Collider.
\end{abstract}

\maketitle

A thorough understanding of the production of an electroweak gauge
boson in association with multiple jets is central to the experimental
physics program at the Large Hadron Collider (LHC).  Such events are
abundant and constitute an important background to many new physics
searches~\cite{ATLAS:2011ad,*Aad:2011qa,*Aad:2011ib,*Aad:2011cwa,
  *Chatrchyan:2011nd,*Chatrchyan:2011ida,*Chatrchyan:2011qs}.  
They typically involve multiple kinematic scales and exhibit polarization 
phenomena~\cite{Bern:2011ie}.
Their study is vital to improve the understanding of quantum chromodynamics
(QCD) in hadron-collider environments~\cite{Aad:2010pg,*Aad:2011xn,
  *Chatrchyan:2011ne,Aad:2012en}, to measure the collider
luminosity, to determine the jet energy scale and to study multiple
parton scattering processes~\cite{Bartalini:2011jp}.  
The most copious event rates occur when the gauge boson is a $W$.  
Predicting $W$+jet production with the most precise theoretical tools 
is therefore of paramount importance for the continued success 
of the LHC physics program.

Typically, good agreement is found when comparing $W$+jets experimental
data with perturbative calculations performed at next-to-leading order
(NLO) in QCD. Corresponding theoretical predictions have recently been made for
associated production with three and even four jets~\cite{KeithEllis:2009bu,*Ellis:2009zw,
  Berger:2008sj,Berger:2009ep,*Berger:2009zg,*Berger:2010zx}.  Despite
their continued success, such calculations suffer from logarithmic
corrections due to intra-jet parton evolution, and from the fact that
they are performed at the parton level.  The latter makes them
unsuitable for direct use in a detector simulation, and requires
additional nonperturbative corrections before a comparison to data can
be performed.  The \MCatNLO~\cite{Frixione:2002ik} and 
\POWHEG~\cite{Nason:2004rx,*Frixione:2007vw} methods remedy this
situation by matching NLO QCD matrix elements with the resummation
encoded in the parton showers of general-purpose Monte-Carlo event
generators~\cite{Buckley:2011ms}, allowing one to obtain well-understood 
hadron-level results at NLO accuracy.  Results obtained with these 
two techniques include $Z$+1-jet production~\cite{Alioli:2010xd}, $W$+2-jets 
production~\cite{Frederix:2011ig} and dijet production~\cite{Alioli:2010xa}.

In this letter we present a new automated approach to matching NLO
results to parton showers.  Our method is a variant of the \MCatNLO
algorithm~\cite{Hoeche:2011fd}  but handles the soft behavior of matrix elements
exactly, for processes with arbitrarily complex color configurations.
We have validated the method for the $W$+jets processes.  Using this
technique, it is now possible, for the first time, to perform a matching
of matrix elements and parton showers for W production in association
with up to three jets at NLO accuracy. This process includes the most
general color topologies, allowing us to demonstrate that the approach
is universal, and permitting its future extension to other processes
with similar or even higher final-state multiplicities.

Our new scheme to implement the \MCatNLO technique is based on the exact 
exponentiation of Catani-Seymour dipole subtraction terms~\cite{Hoeche:2011fd}.
This method allows to circumvent the otherwise occurring integral over residual 
real-radiative contributions to the NLO cross section, that arise from the 
modified subtraction scheme in \MCatNLO~\cite{Frixione:2002ik}.  It also 
allows, for the first time, to obtain the correct soft-gluon limit in the 
first emission of the parton shower, such that no ad-hoc adjustments to the 
splitting kernels must be made.
In fact our approach can be shown to correctly take into account the full 
color structure of the processes at NLO and in particular to correctly
reproduce the soft gluon limit.

The \MCatNLO cross section can be written as~\cite{Frixione:2002ik,Hoeche:2011fd}
\begin{equation}
  \label{eq:mcatnlo_xs}
  \begin{split}
    \sigma\,=&\;\int\done\Phi_B\,\bar{\mr{B}}^{\rm(A)}(\Phi_B)\,
    \bigg[\,\bar{\Delta}^{\rm(A)}(t_0)\\
      &\quad\qquad+\int_{t_0}\done\Phi_1
      \frac{\mr{D}^{\rm(A)}(\Phi_B,\Phi_1)}{\mr{B}(\Phi_B)}\,
      \bar{\Delta}^{\rm(A)}(t)\,\bigg]\\
    &\quad+\int\done\Phi_R\,\mr{H}^{\rm(A)}(\Phi_R)\;,
  \end{split}
\end{equation} 
where
\begin{equation}
  \label{eq:mcatnlo_bbar}
  \begin{split}
    &\bar{\mr{B}}^{\rm(A)}(\Phi_B) \,=\;
    \mr{B}(\Phi_B)+\tilde{\mr{V}}(\Phi_B)+\mr{I}^{\rm(S)}(\Phi_B)\\
    &\quad+\int\done\Phi_1\left[\vphantom{\int}\,
      \mr{D}^{\rm(A)}(\Phi_B,\Phi_1)-\mr{D}^{\rm(S)}(\Phi_B,\Phi_1)\,\right]
  \end{split}
\end{equation}
The terms $\mr{B}$, $\tilde{\mr{V}}$, $\mr{I}^{\rm(S)}$, and $\mr{D}^{\rm(S)}$ 
represent the Born contribution, virtual correction plus collinear 
counterterms, integrated subtraction terms and real subtraction terms. 
$\Phi_B$ and $\Phi_R$ denote Born- and real-emission phase space with 
$\Phi_R=\Phi_B\otimes\Phi_1$, where $\Phi_1$ represents the phase space of the
respective additional parton emission.  Real-emission matrix elements $\mr{R}$ 
are separated into an infrared-singular (soft) and an infrared-regular (hard) 
part, $\mr{D}^{\rm(A)}$ and $\mr{H}^{\rm(A)}$, where 
$\mr{R} = \mr{D}^{\rm(A)} + \mr{H}^{\rm(A)}$.
This leads to the definition of the Sudakov form factor
\begin{equation}
  \label{eq:mcatnlo_sudakov}
  \bar{\Delta}^{\rm(A)}(t,t') \,=\; 
  \exp\cbr{-\int_t^{t'}\done\Phi_1\,
  \frac{\mr{D}^{\rm(A)}(\Phi_B,\Phi_1)}{\mr{B}(\Phi_B)}}\,.
\end{equation} 
The key point of our new technique is that the integral in 
Eq.~\eqref{eq:mcatnlo_bbar} is avoided, since in our approach 
$\mr{D}^{\rm(A)}=\mr{D}^{\rm(S)}$, i.e.\ the subtraction kernels are also employed
for parton showering. This can be achieved using Catani-Seymour subtraction,
by dynamically correcting a parton shower based on spin-and color-averaged 
splitting operators. The method was applied previously to the $W^\pm$- $Z$- 
and Higgs+1-jet production processes~\cite{Hoeche:2011fd}.  In this 
publication we show that it is not limited to the case of one final-state 
parton at Born level, with a relatively trivial colour structure.
We present results for $W$+2- and $W$+3-jet production, which contain the 
most general non-trivial color structures.

\begin{table*}[thb]
  \renewcommand{\arraystretch}{1.5}
  \addtolength{\tabcolsep}{5pt}
  \begin{center}
    \begin{tabular}{|c|c|c|c|c|}
      \hline
      $W^\pm+\geq n$ jets &  \ATLAS  &  NLO  &  \MCatNLO 1em  &  \MCatNLO PL \\
      \hline
      \hline
      $n=0$ & 5.2$\,\pm\,$0.2 & 5.06(1) & 5.09(3) & 5.06(3) \\
      \hline
      $n=1$, $p_{\perp\,j}>20$ GeV & 0.95$\,\pm\,$0.10 & 0.958(5) & 0.968(10) & 0.889(10) \\[-2pt]
      \hphantom{$n=1$,} $p_{\perp\,j}>30$ GeV & 0.54$\,\pm\,$0.05 & 0.527(4) & 0.534(7) & 0.474(7) \\
      \hline
      $n=2$, $p_{\perp\,j}>20$ GeV & 0.26$\,\pm\,$0.04 & 0.263(2) & 0.260(5) & 0.236(4) \\[-2pt]
      \hphantom{$n=2$,} $p_{\perp\,j}>30$ GeV & 0.12$\,\pm\,$0.02 & 0.120(1) & 0.123(2) & 0.109(2) \\
      \hline
      $n=3$, $p_{\perp\,j}>20$ GeV & 0.068$\,\pm\,$0.014 & 0.072(3) & 0.059(3) & 0.060(3) \\[-2pt]
      \hphantom{$n=3$,} $p_{\perp\,j}>30$ GeV & 0.026$\,\pm\,$0.005 & 0.026(1) & 0.022(2) & 0.021(1) \\
      \hline
  \end{tabular}
  \end{center}
  \caption{Total cross sections in nb for $W^\pm+\geq 0,1,2,3$ jet production as measured by
    \ATLAS~\cite{Aad:2012en} compared to predictions from the
    corresponding fixed order calculations, and matrix-element/shower level
    \MCatNLO simulations. Statistical uncertainties of the theoretical predictions
    are quoted in parentheses.
  }
  \label{tab:jetmulti}
\end{table*}


We use the \Sherpa event generator~\cite{Gleisberg:2003xi,*Gleisberg:2008ta},
including its automated \MCatNLO implementation~\cite{Hoeche:2011fd}.  The 
finite part of virtual corrections is computed using the \BlackHat 
library~\cite{Berger:2008sj}, the Born part and phase space integration is
provided by the matrix-element generator \Amegic~\cite{Krauss:2001iv}, 
including an automated implementation~\cite{Gleisberg:2007md} of the 
Catani-Seymour dipole subtraction method~\cite{Catani:1996vz}.  The parton 
shower model~\cite{Schumann:2007mg,*Hoeche:2009xc} uses transverse momentum 
as ordering parameter, thus avoiding the problem of truncated 
emissions~\cite{Nason:2004rx,*Frixione:2007vw}.  We restrict the resummation
region using the methods described in~\cite{Hoeche:2012fm} by setting the resummation
scale, $\mu_Q$, identical to the factorization scale, $\mu_F$.
We analyze the dependence of our results on the resummation scale
by varying it in the range $\sqrt{1/2}\mu_Q\ldots \sqrt{2}\mu_Q$. We compare
this variation with the uncertainty of the NLO calculation that arises from
varying renormalization and factorization scales in the range 
$1/2\mu_{R/F}\ldots 2\mu_{R/F}$~\cite{Berger:2010zx}.

Note that for the $W$+3-jet virtual matrix element we use the leading-color 
approximation in \BlackHat only, to avoid an unnecessary increase in CPU time 
for the simulation. Subleading color configurations in virtual corrections 
often play a minor role in $W$+multi-jet processes~\cite{Ita:2011ar}.  They 
might, however, be important in other situations. As we focus on the interface 
between the NLO calculation and the parton shower in fairly inclusive 
observables, sub-leading colour effects are neglected.  The CTEQ6.6 PDF 
set~\cite{Nadolsky:2008zw} is employed together with the corresponding 
parametrization of the running coupling. Following~\cite{Berger:2010vm}
renormalization and factorization scales are chosen as 
$\mu_R=\mu_F=1/2\,\hat{H}_T'$, where $\hat{H}_T'=\sqrt{\sum p_{T,j}^2+E_{T,W}^2}$. 
Predictions are presented at two different levels of event simulation:
\begin{description}
\item[``NLO''] Fixed-order matrix-element calculation,
\item[``MC@NLO PL''] \MCatNLO including full parton showering, but no non-perturbative effects.
\end{description}
The aim of this study is to present and validate an application of the 
\MCatNLO variant suggested in~\cite{Hoeche:2011fd} to processes with
complex QCD final states.  Therefore, non-perturbative effects, stemming
from hadronization, hadron decays or multiple parton interactions are 
neglected in this study\footnote{The observables displayed here are 
relatively insensitive to non-perturbative corrections and have been analyzed 
in detail in~\cite{Hoeche:2011fd}.}.  The scale uncertainties of NLO results
are quoted to gauge the resummation uncertainties from the \MCatNLO.

The analysis is carried out with the help of Rivet~\cite{Buckley:2010ar}
following a recent study of $W^\pm$+jets production by the \ATLAS
collaboration~\cite{Aad:2012en}. Events are selected to contain a 
lepton within $|\eta|<2.5$ with $p_\perp>20$~GeV and requiring
$E_T^\text{miss}>25$~GeV. A cut on $m_\mathrm{T}^{\mathrm{W}}>40$~GeV is 
additionally applied.  All particles other than the leading electron and 
neutrino are clustered into anti-$k_t$ jets with $R=0.4$ and $p_\perp>30$~GeV. 
The analysis is carried out in jet multiplicity bins up to $N=3$
and cross sections are studied differentially in several observables.

The results for each observable are predicted at NLO accuracy,
i.~e.~all differential cross sections for $W^\pm+\geq n$-jet events are
generated using the $W^\pm+n$-jet NLO or \MCatNLO calculation.
For $n>0$, the $W$+n-jet matrix element must be regularized by requiring at 
least $n$ jets with a minimum transverse momentum. This cut is chosen 
to be $p_\perp^\text{gen}>10$~GeV to make the event sample inclusive enough
for the analysis. We have checked that our results are independent of the
precise value of this cut by varying it from 5 to 15~GeV in every 
individual jet bin.

Table \ref{tab:jetmulti} compares total cross sections in four inclusive jet
multiplicity bins. The \ATLAS measurement is reproduced very well both by the
fixed order calculation as well as by the \MCatNLO matched simulation.
The agreement between the NLO results and the \MCatNLO simulation is excellent,
indicating that the matching to the parton shower does not alter the jet 
production rate as predicted by the fixed-order calculation.

\begin{figure}[thb]
  \centering
  \includegraphics[width=\linewidth]{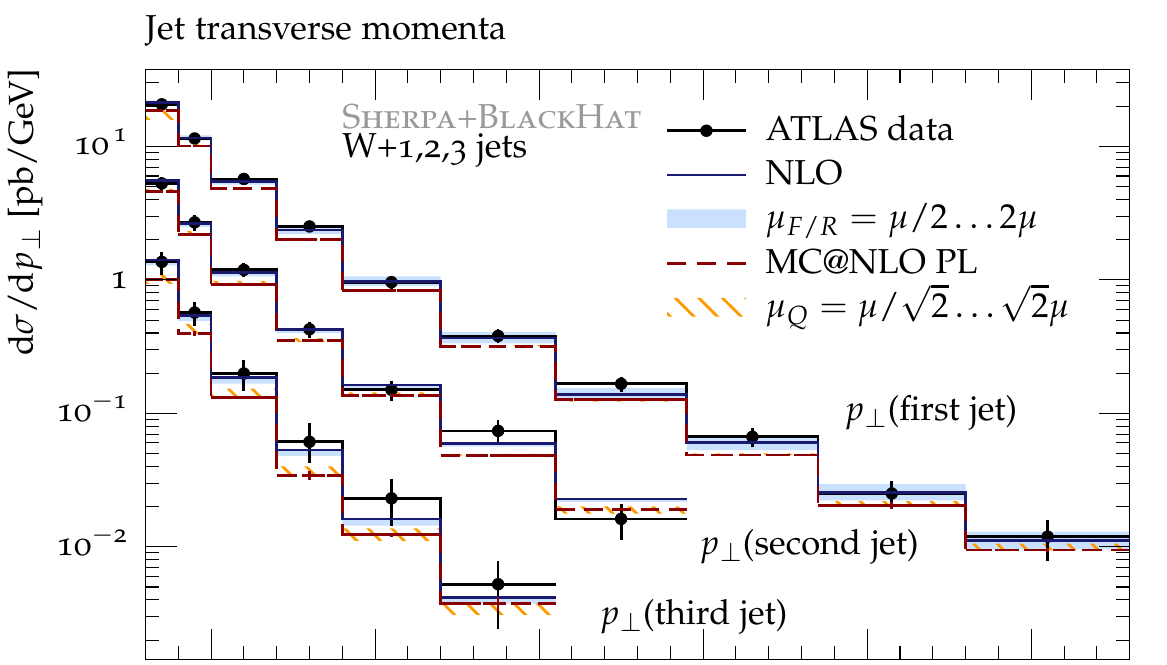}
  \includegraphics[width=\linewidth]{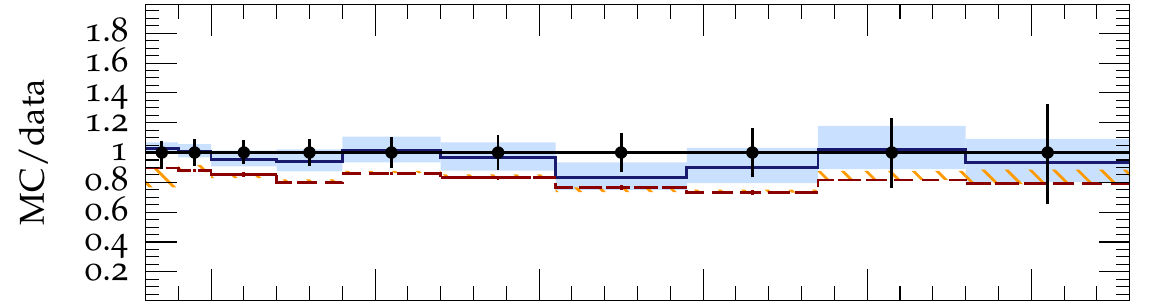}
  \includegraphics[width=\linewidth]{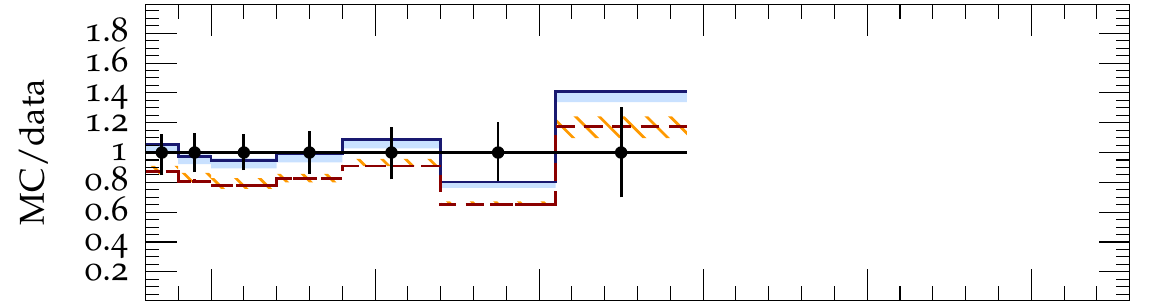}
  \includegraphics[width=\linewidth]{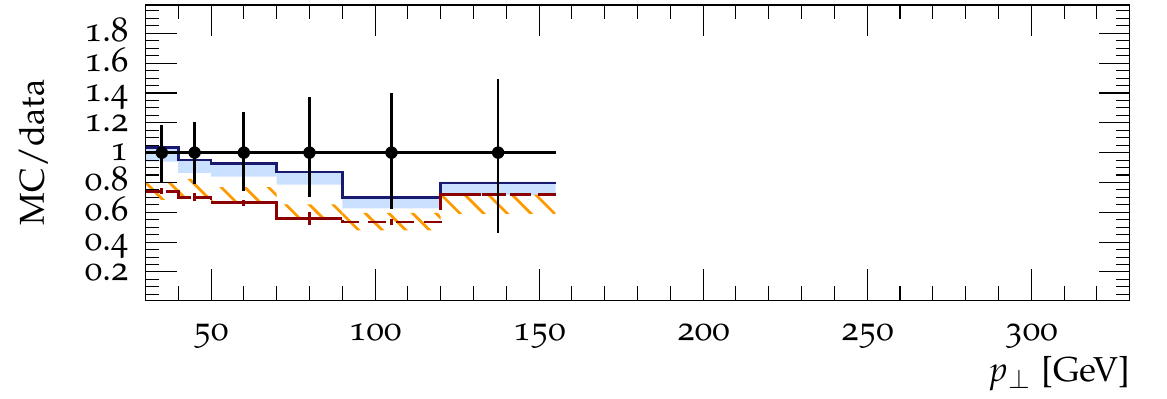}
  \caption{Transverse momentum of the first, second and third jet (from top
    to bottom) in $W^\pm+\geq 1,2,3$ jet production as measured by
    \ATLAS~\cite{Aad:2012en} compared to predictions from the
    corresponding fixed order and \MCatNLO simulations. The blue band display
    fixed-order uncertainties, the orange band shows resummation uncertainties.
  }
  \label{fig:jetpt}
\end{figure}

In Fig.~\ref{fig:jetpt} we display a comparison of the transverse momentum
spectra of the first, second and third hardest jet in $W+\geq 1$-, 2- and 
3-jet production. No significant changes are observed when switching from
the fixed-order calculation to the \MCatNLO simulation, again indicating 
that the hard kinematics predicted by the NLO result are respected in the 
subsequent parton-shower evolution.

\begin{figure}[thb]
  \centering
  \includegraphics[width=\linewidth]{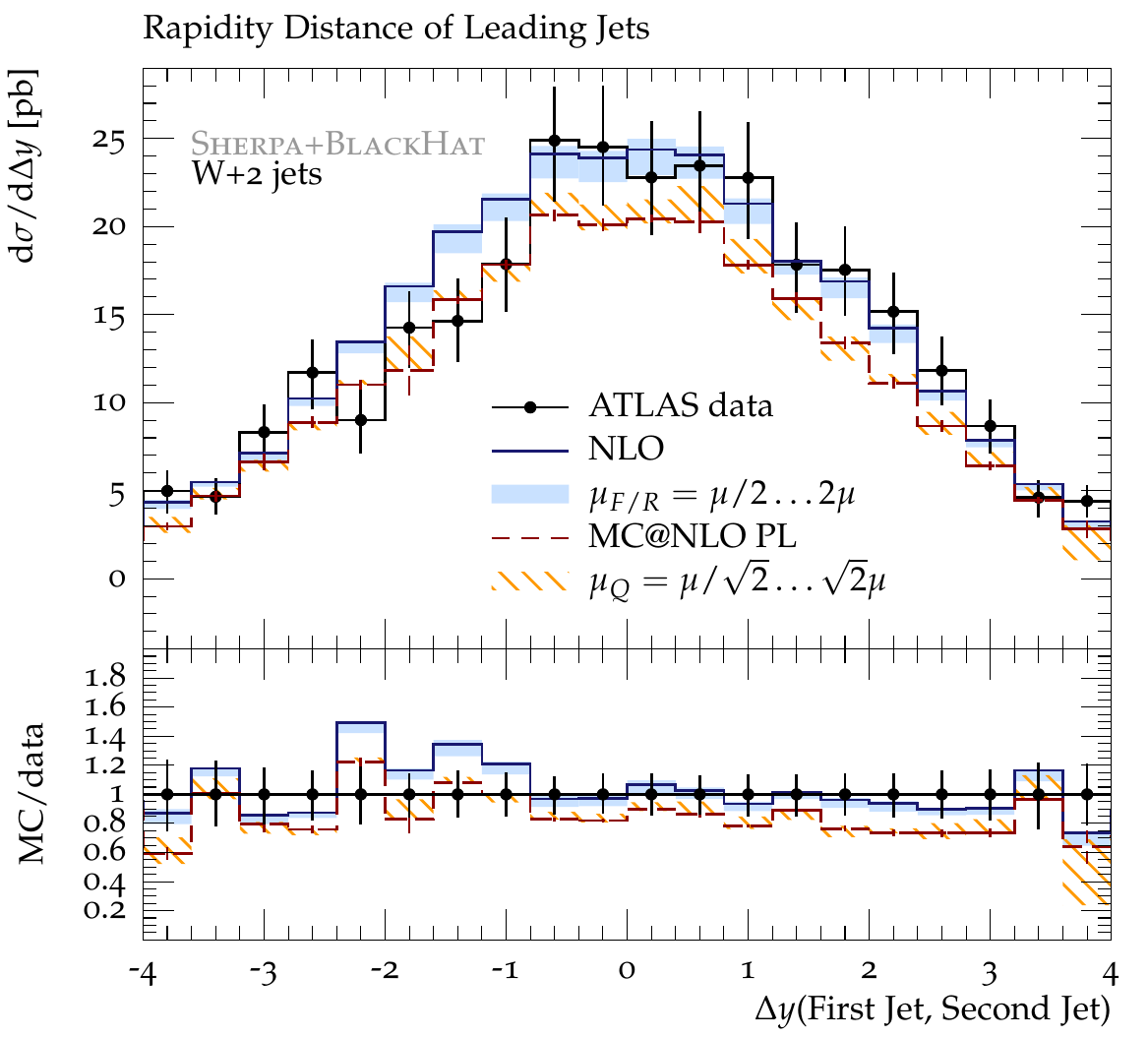}
  \includegraphics[width=\linewidth]{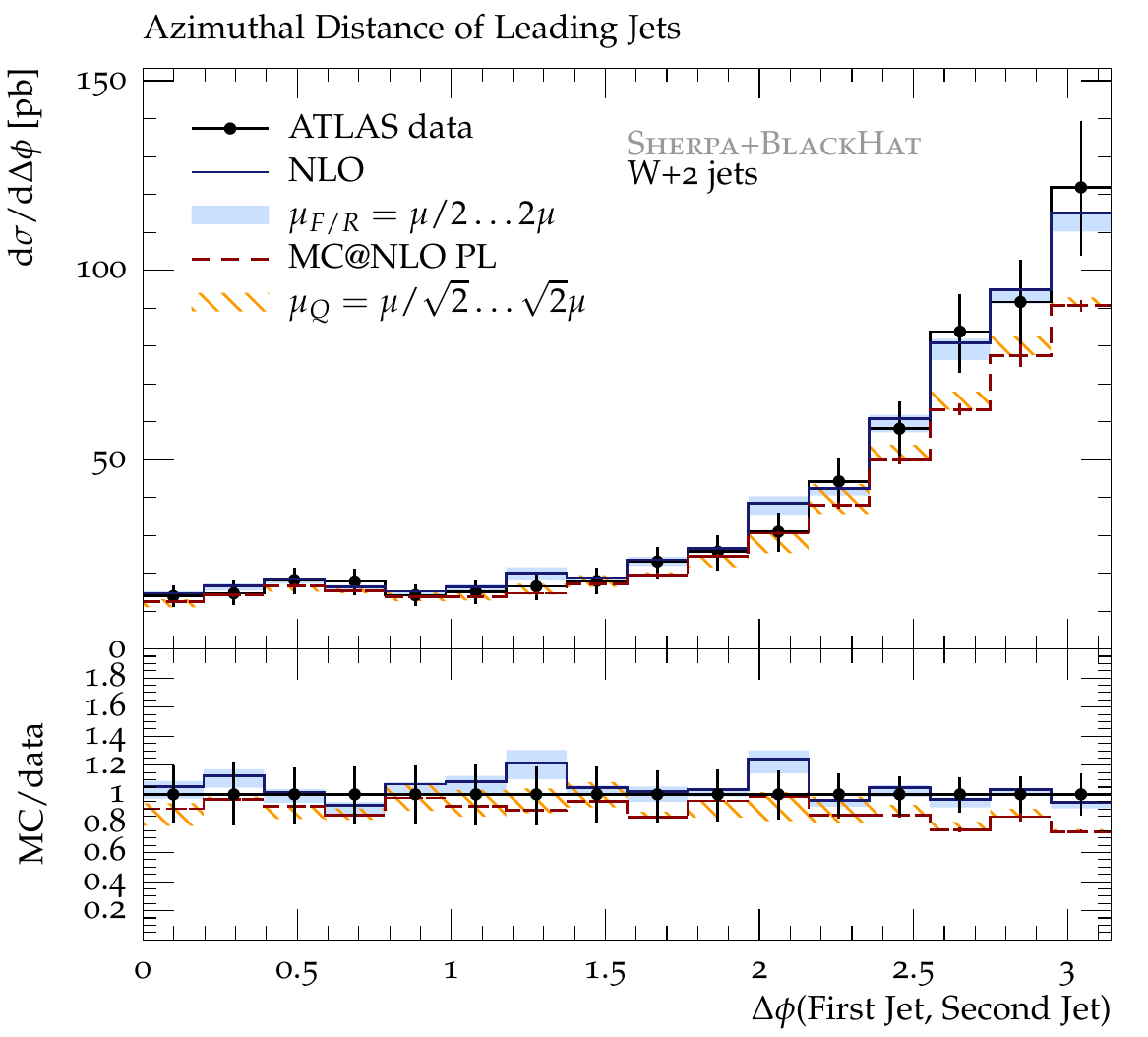}
  \caption{Angular correlations of the two leading jets in
    $W^\pm+\geq 2$ jet production as measured by
    \ATLAS~\cite{Aad:2012en} compared to predictions from the
    $W^\pm+2$ jet fixed order and \MCatNLO simulations. The blue band display
    fixed-order uncertainties, the orange band shows resummation uncertainties.
  }
  \label{fig:dijets}
\end{figure}

Fig.~\ref{fig:dijets} focuses on $W+\geq 2$-jet events. Angular correlations
between the two leading jets are sensitive to QCD corrections in the $W+2$-jet
process and are thus a useful observable to validate the QCD radiation pattern
which is generated in our \MCatNLO. Both, the rapidity and azimuthal separation 
of the jets are predicted in perfect agreement with data.

In summary, we have shown in this letter how our recently proposed method 
for implementing \MCatNLO can be used to produce novel and relevant results 
for one of the most challenging collider signatures to date.  We have 
compared results for $W$+0-, 1-, 2- and 3-jet production to recent \ATLAS 
data and found excellent agreement for all observables, with only a selection 
of them presented here.  In so doing, for the first time results for $W$+3-jets
production were presented.  The success and the simplicity of our \MCatNLO 
variant make it a prime candidate for the implementation of a matrix-element 
parton-shower merging algorithm at next-to-leading order.

\bigskip

\begin{acknowledgments}
We would like to thank Zvi Bern, Stefan Dittmaier, Lance Dixon, Harald Ita,
David Kosower and Kemal Ozeren for many fruitful discussions. We are especially
grateful to Lance Dixon for carefully reading the manuscript.

SH's work was supported by the US Department of Energy under contract 
DE--AC02--76SF00515, and in part by the US National Science Foundation, grant 
NSF--PHY--0705682, (The LHC Theory Initiative).  MS's work was supported by 
the Research Executive Agency (REA) of the European Union under the Grant 
Agreement number PITN-GA-2010-264564 (LHCPhenoNet). FS's work was supported
by the German Research Foundation (DFG) via grant DI 784/2-1.
We gratefully thank the bwGRiD project~\cite{bwgrid} for the computational resources.
\end{acknowledgments}

\bibliography{journal}
\end{document}

%% file: defs.tex
\usepackage{amsmath}
\usepackage{amssymb}
\usepackage{array}
\usepackage{pstricks}
\usepackage{graphicx}
\usepackage{xspace}
\makeatletter
\DeclareRobustCommand*{\bfseries}{%
  \not@math@alphabet\bfseries\mathbf
  \fontseries\bfdefault\selectfont
  \boldmath
}
\makeatother
\newcommand{\MCatNLO}{M\protect\scalebox{0.8}{C}@N\protect\scalebox{0.8}{LO}\xspace}

\newcommand{\POWHEG}{P\protect\scalebox{0.8}{OWHEG}\xspace}

\newcommand{\BlackHat}{B\protect\scalebox{0.8}{LACK}H\protect\scalebox{0.8}{AT}\xspace}

\newcommand{\Sherpa}{S\protect\scalebox{0.8}{HERPA}\xspace}

\newcommand{\Amegic}{A\protect\scalebox{0.8}{MEGIC++}\xspace}

\newcommand{\ATLAS}{ATLAS\xspace}

\long\def\symbolfootnote[#1]#2{\begingroup%
\def\thefootnote{\fnsymbol{footnote}}\footnote[#1]{#2}\endgroup}

\newcommand{\cbr}[1]{\left\{ #1\right\}}

\newcommand{\done}{{\rm d}}

\newcommand{\mr}[1]{\mathrm{#1}}
